\documentclass[prl,twocolumn]{revtex4}
\usepackage{natbib}
\usepackage[latin9]{inputenc}
\usepackage{units}
\usepackage{amsmath}
\usepackage{amssymb}
\usepackage{graphicx}
\usepackage{esint}
\usepackage{braket}
\usepackage{diagbox}

\makeatletter
\makeatletter

\newcommand{\Yb}{\ensuremath{^{171}\mathrm{Yb}^+~}}
\newcommand{\up}{\ensuremath{\left|\uparrow\right\rangle}}
\newcommand{\down}{\ensuremath{\left|\downarrow\right\rangle}}

\makeatother

\begin{document}
\title{Realization of near-deterministic arithmetic operations and quantum state engineering}
\author{Mark Um$^{1}$, Junhua Zhang$^{1}$, Dingshun Lv$^{1}$, Yao Lu$^{1}$, Shuoming An$^{1}$, Jing-Ning Zhang$^{1}$, Hyunchul Nha$^{2,3}$, M. S. Kim$^{4*}$ and Kihwan Kim$^{1\dagger}$}

\affiliation{$^{1}$Center for Quantum Information, Institute for Interdisciplinary Information Sciences, Tsinghua University, Beijing 100084, P. R. China  \\ $^{2}$School of Computational Sciences, Korea Institute for Advanced Study, Seoul 130-722, Korea\\ $^{3}$Texas A$\&$M University at Qatar, Education City, P.O. Box 23874, Doha, Qatar \\$^{4}$QOLS, Blackett Laboratory, Imperial College London, SW7 2AZ, United Kingdom} 







\begin{abstract}
Quantum theory is based on a mathematical structure totally different from conventional arithmetic. Due to the symmetric nature of bosonic particles, annihilation or creation of single particles translates a quantum state depending on how many bosons are already in the given quantum system. This proportionality results in a variety of non-classical features of quantum mechanics including the bosonic commutation relation. The annihilation and creation operations have recently been implemented in photonic systems. However, this feature of quantum mechanics does not preclude the possibility of realizing conventional arithmetic in quantum systems. We implement conventional addition and subtraction of single phonons for a trapped \Yb ion in a harmonic potential. In order to realize such operations, we apply the transitionless adiabatic passage scheme on the anti-Jaynes-Cummings coupling between the internal energy states and external motion states of the ion. By performing the operations on superpositions of Fock states, we realize the hybrid computation of classical arithmetic in quantum parallelism, and show that our operations are useful to engineer quantum states. Our single-phonon operations are nearly deterministic and robust against parameter changes, enabling handy repetition of the operations independently from the initial state of the atomic motion. We demonstrate the transform of a classical state to a nonclassical one of highly sub-Poissonian phonon statistics and a Gaussian state to a non-Gaussian state, by applying a sequence of the operations. The operations implemented here are the Susskind-Glogower phase operators, whose non-commutativity is also demonstrated. 
\end{abstract}

\maketitle
\section{I. Introduction}
Due to the symmetric nature of bosons \cite{Bose1924}, the probability of creating or annihilating a boson is proportional to the number of bosons in the original state \cite{DiracQM}. The single boson operations thus bear totally different mathematical structure from conventional arithmetic of addition and subtraction, which causes the predictions of quantum mechanics different from those of classical mechanics and provides a foundational basis for commutation relations. Recently, there have been seminal works to realize the bosonic operations at the single-boson level for the test of foundations and applications of quantum mechanics \cite{Grangier06,Polzik06,Bellini07,Bellini08,Bellini09,Inoue10,Lvovsky13}. Here, however, we address a fundamental question, 'Does the existence of bosonic annihilation and creation exclude the possibility of realizing the conventional arithmetic subtraction and addition operations in quantum mechanics?' In our work, we answer this question by realizing such operations on the phonon states of a trapped ion in a harmonic potential. We perform these operations on coherent states and superpositions of Fock states, demonstrating a hybrid of conventional arithmetic and quantum parallelism. The seemingly simple operations which match to classical concepts of subtraction and addition bring a classical state into a non-classical state nearly deterministically. The technology we demonstrate here can be used in many quantum systems including circuit quantum electrodynamics\cite{Joo15} to enable quantum state engineering. 

The conventional addition and subtraction of a particle can be written as
\begin{eqnarray}
\hat{S}^{+}=\sum_{n=0} \ket{n+1}\bra{n}, \hat{S}^{-}=\sum_{n=1} \ket{n-1}\bra{n}.
\label{eq:addsubtract}
\end{eqnarray}
where $\ket{n}$ stands for a Fock state of $n$ bosons. $\hat{S}^+$ takes the $n$-particle state to the $(n+1)$-state representing an addition, while the subtraction operation, $\hat{S}^-$, brings the $n$-state to the $(n-1)$-state. These operations correspond to conventional arithmetic which is commonly used in everyday life but they do not come out naturally in quantum mechanics. Instead, in quantum-mechanics, creation $\hat{a}^\dag$ and annihilation $\hat{a}$ operators are introduced, which bear the operator relations  
\begin{equation}
\hat{a}^\dag=\sum_{n=0} \sqrt{n+1}\ket{n+1}\bra{n}~~,~~~ \hat{a}=\sum_{n=0}\sqrt{n} \ket{n-1}\bra{n}.
\label{boson}
\end{equation}
The proportionality factors $\sqrt{n+1}$ and $\sqrt{n}$ appear due to the symmetric indistinguishable nature of bosons \cite{DiracQM}. The addition or subtraction in quantum domain thus involves the modification of the probability amplitude of state due to the excitation $n$-dependent factor. While the quantum operators (\ref{boson}) have been experimentally demonstrated \cite{Grangier06,Polzik06,Bellini07,Bellini08,Bellini09,Inoue10,Lvovsky13}, the realization of the conventional operations (\ref{eq:addsubtract}) is still to be attested in the quantum regime. Indeed, the $\hat{S}^+$ and $\hat{S}^-$ operators were suggested as the elements of a phase operator by Susskind and Glogower \cite{Glogower1964}. Thus realization of such operations would serve as an important stepping stone to study the properties of the Susskind-Glogower phase operator experimentally. In this paper, we demonstrate the operations in a near deterministic manner, also showing that the technique is a resource to create non-Gaussian states efficiently \cite{Jeffers13,Wilhelm14}. We show that classical states are turned into nonclassical ones manifesting highly sub-Poissonian photon statistics and negativity in the Wigner function. The versatility of the operations for quantum state engineering is demonstrated by various sequences of the single-phonon operations. This is contrasted to the bosonic operations realized so far \cite{Grangier06,Polzik06,Bellini07,Bellini08,Bellini09,Inoue10,Lvovsky13}. Their success probability is intrinsically low (indeed, the higher the fidelity of the operations, the lower the success rate); hence a repetition of such the operations is practically limited.

\begin{figure}[hp]
\includegraphics[width=1\columnwidth]{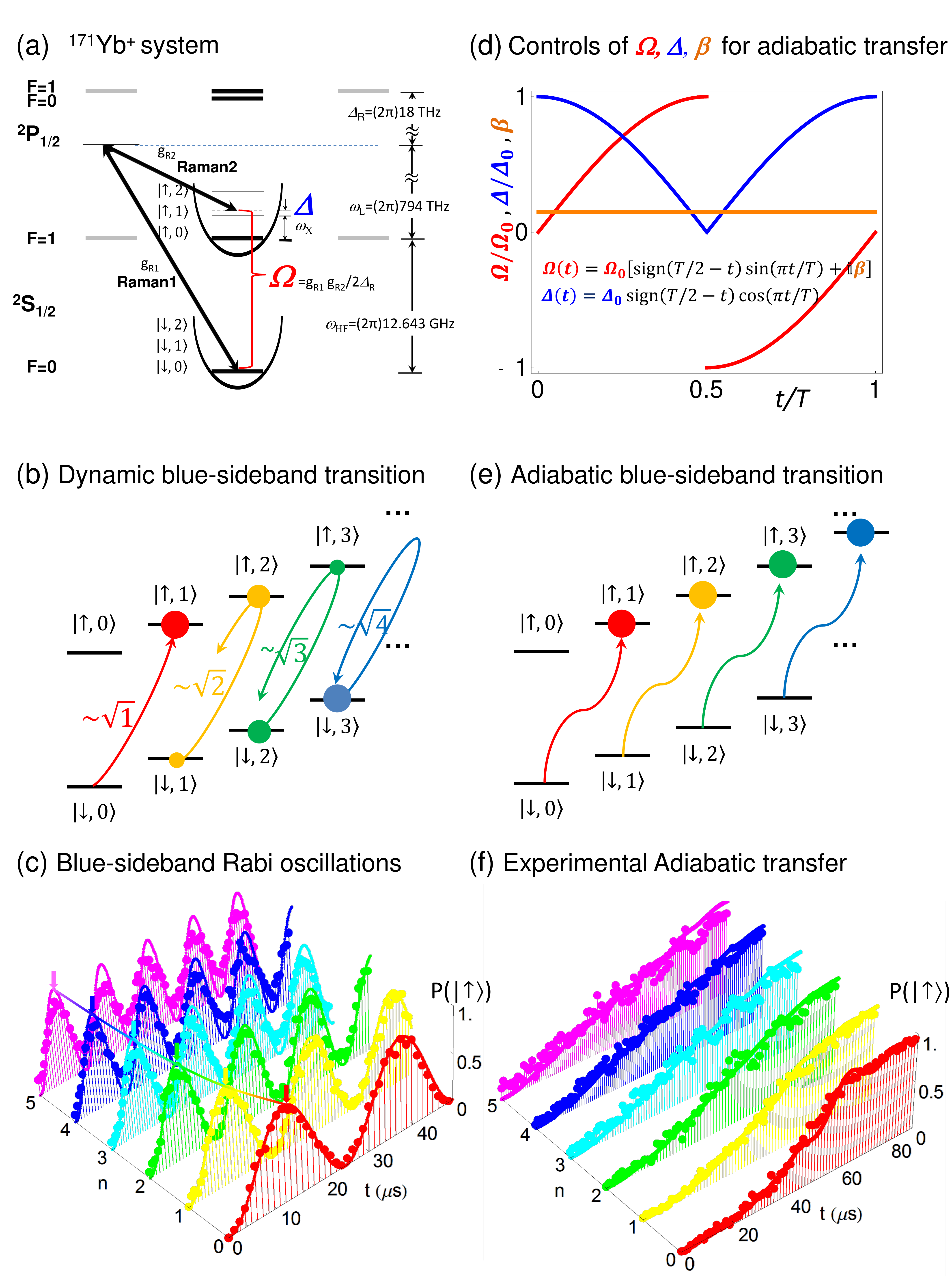}
\caption{Experimental scheme and dynamic and adiabatic transition by anti-Jaynes-Cummings interaction. (a) The \Yb system in a harmonic potential. The qubit level in $S_{1/2}$ manifold, $\ket{F=0, m_{F}=0} \equiv$ \down$~$ and $\ket{F=1, m_{F}=0} \equiv$ \up$~$ are coupled by the Raman laser beams to produce the anti-Jaynes-Cummings interaction or blue-sideband transition. We denote $\Omega$ as the Rabi-frequency on the qubit transition and the $\Delta$ is the frequency difference of the beat-note frequency of Raman beams from $\omega_{\rm HF}+\omega_{\rm X}$. (b) The diagram for the standard blue-sideband transition. The Hilbert space is composed of the qubit states, \up and \down, and phonon states of $n$ excitations, $\ket{n}$. The transitions between $\ket{\downarrow, n}$ and $\ket{\uparrow, n+1}$ would experience evolutions $\sqrt{n+1}$ times faster than the transition between $\ket{\downarrow, 0}$ to $\ket{\uparrow, 1}$. (c) Probability of finding the ion in $|\uparrow, n\rangle$ state as a function of time, which clearly manifests $\sqrt{n+1}$-dependence. The arrow at the first peak of each oscillation indicates the duration of a $\pi$-pulse for the corresponding transition. The $\pi$-pulse duration, $T_{\pi}$, of the fundamental blue-sideband transition (red) is $13~\mu$s. The dots represent experimental data and solid lines are from the fitting to $\sin^2 \left(\frac{\sqrt{n+1} \pi}{2 T_{\pi}} t \right)$. (d) For the adiabatic blue-sideband transition whose frequency is independent of motional quantum number $n$. The $\Omega$ and $\Delta$ are controlled as the red and blue curves. The $i \beta$ in $\Omega$ is the counter-diabatic term to suppress the transition during the evolution. Here $\Omega_{0} = (2 \pi) 38.5$ kHz, $\beta = 0.075$, and $\Delta_{0} = 1.6 \Omega_{0}$. (e) The conceptual diagram for the adiabatic blue-sideband transition without the $\sqrt{n+1}$-dependence. (f) The experimental demonstration of the adiabatic blue-sideband transitions realized by the transitionless quantum driving. The total time to execute the transitions is $91 \mu$s for any $\ket{n}$, which is about 7 times $T_{\pi}$.}
\label{fig1:concept}
\end{figure}

\begin{figure*}[ht]
\includegraphics[width=2.\columnwidth]{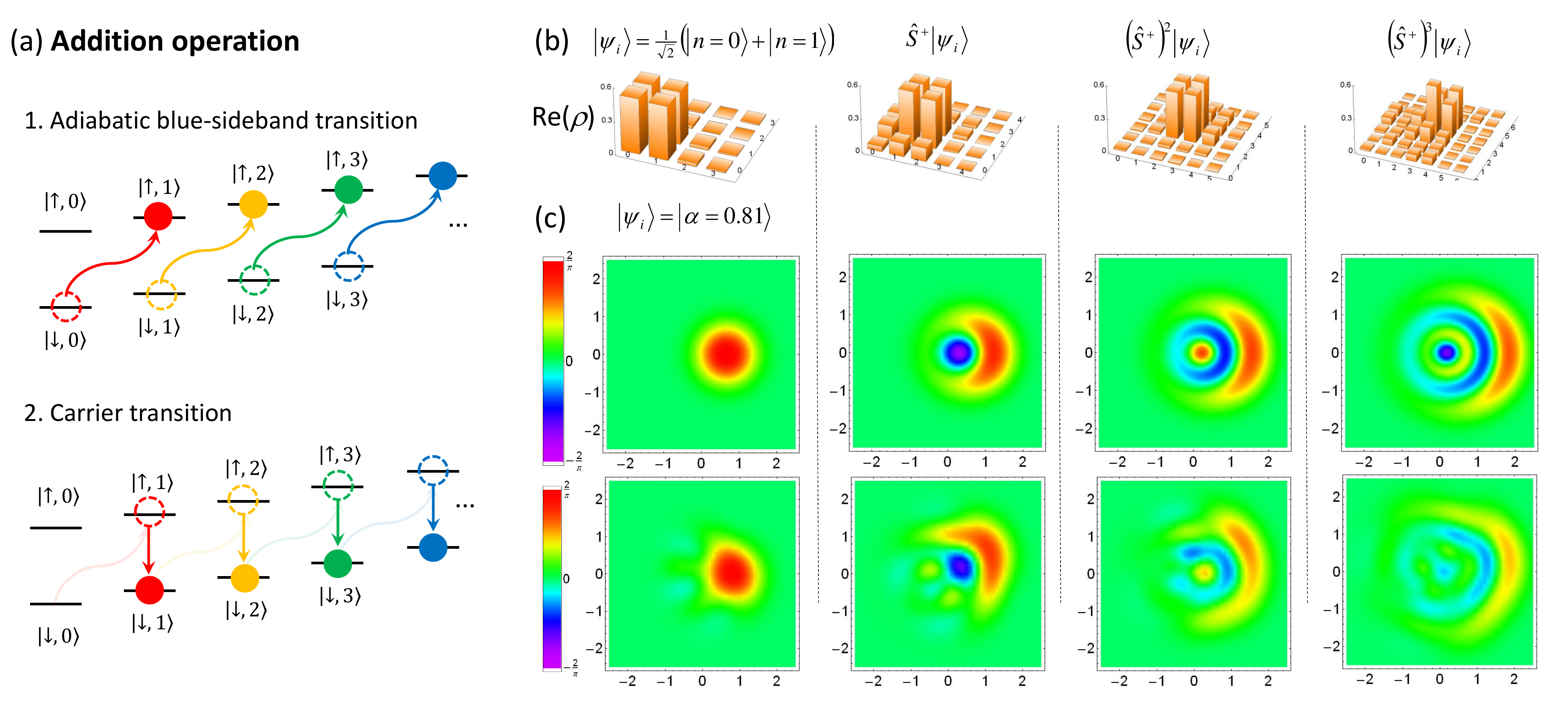} 
\caption{Schematic diagram and experimental results for the phonon addition. (a) Implementation of addition is composed of a $\pi$-pulse of adiabatic blue-sideband transition which transfers all the phonons from $\ket{\downarrow, n}$ to $\ket{\uparrow, n+1}$ uniformly, followed by a $\pi$-pulse of carrier transition which couples the internal states resonantly, bringing $\ket{\uparrow, n}$ down to $\ket{\downarrow, n}$. This realizes the addition operation $\hat{S}^{+}$ of Eq. (\ref{eq:addsubtract}). (c) Additions on a superposition state $\ket{\psi_{i}}=\frac{1}{\sqrt{2}}\left( \ket{n=0}+\ket{n=1}\right)$. The reconstructed density matrices, only the real part of them, indicate the fidelity 0.99 ($<0.01$) of the initially prepared state and those of the final states 0.96(0.01), 0.92(0.01) and  0.87(0.01) after one, two and three times addition, respectively. The purities of the output states are 0.92(0.01),  0.81(0.03), and  0.71(0.06), respectively. The numbers in the parentheses represent the sizes of error estimated by the maximal-likelihood methods (see supplementary information). This clearly shows the capability of keeping coherence. (c) Wigner functions of the $n$ phonon-added coherent state of amplitude $\alpha=0.8$ where $n=0,1,2$ and 3. Observed negative values in the Wigner function proves the production of non-Gaussian state which results from a simple operation of conventional addition. The fidelities are reduced from 0.97(0.01) for the initial state to 0.87(0.01) (one single-phonon addition), 0.84(0.01)(two additions), 0.85(0.02) (three additions) and purities are changed from 0.99 to 0.93(0.02), 0.93(0.02), 0.80(0.03).}
\label{fig2:addition}
\end{figure*}

\section{II. Experimental Implementation}
We implement the $\hat{S}^+$ and $\hat{S}^-$ operations on a motional mode of frequency $\omega_{\rm X}$ for a single trapped \Yb ion in a three-dimensional harmonic potential \cite{Monroe96,DidiRMP,Jost09} as shown in Fig. \ref{fig1:concept}(a). The harmonic potential is generated by a rotating electric field in the radial axis with trap frequencies $\left(\omega_{\rm X},\omega_{\rm Y}\right) = (2 \pi) (2.8,3.2)$ MHz and a dc voltage in the axial direction with $\omega_{\rm Z}= (2 \pi) 0.6$ MHz. The two hyperfine states $\ket{F=1,m_{F}=0}\equiv\ket{\uparrow}$ and $\ket{F=0,m_{F}=0}\equiv\ket{\downarrow}$ of the $S_{1/2}$ manifold  represent a $qubit$ state with their transition frequency $\omega_{\rm HF} = (2 \pi) 12.6428$ GHz. The anti-Jaynes-Cummings interaction, i.e., blue sideband transition, $H_{\rm aJC} = \frac{\eta \Omega}{2} \hat{a}^{\dagger} \hat{\sigma}_{+} e^{i \Delta t}+ {\rm h.c.} $, is realized by the stimulated Raman laser beams with beat-note frequency $(\omega_{\rm R1}-\omega_{\rm R2}) = (\omega_{\rm HF} + \omega_{\rm X})+ \Delta$, where $\Omega$ is the Rabi frequency for the qubit transition, $\eta = \Delta k \sqrt{\hbar/2 M \omega_{\rm X} }$ the Lamb-Dicke parameter, $\Delta k$ the net wave-vector of the Raman laser beams and $M$ the mass of \Yb ion. The configuration of the laser beams produces the transition between $\ket{\downarrow, n}$ and $\ket{\uparrow, n+1}$ with the oscillation frequency of $\sqrt{n+1} \eta \Omega$ as shown in Fig. \ref{fig1:concept}(b), due to the fundamental property of $\hat{a}^{\dagger}$ and $\hat{a}$ operators in Eq. (\ref{boson}). Therefore, it is fundamentally impossible to transfer $\ket{\downarrow, n}$ to $\ket{\uparrow, n+1}$ in an $n$ independent manner by applying the blue-sideband beams at a fixed duration of time. Fig. \ref{fig1:concept}(c) shows the experimental time-evolution of the blue-sideband transition, where the oscillation rates are clearly increased by $\sqrt{n+1}$ factor. 

Here we note that an adiabatic transition shown in Fig. \ref{fig1:concept}(e), such as the stimulated Raman adiabatic transition \cite{Shore98}, will result in removing $\sqrt{n+1}$ and $\sqrt{n}$ in the dynamics and enable us to realize $\hat{S}^+$ and $\hat{S}^-$ operations. We experimentally erase the excitation-dependent factor in the population transfer from the $\ket{\downarrow, n}$ state to the $\ket{\uparrow, n\pm 1}$ state by applying the scheme of transitionless driving \cite{Berry09,Muga10,Morsch12,Junhua14}. Generally an adiabatic method provides a robust transfer from one state to another against serious fluctuation in control parameters of a system. For a perfect adiabatic transfer, however, a relatively slow operation is required, which makes it impractical in many quantum systems. We experimentally reduce the duration of the adiabatic transfer by adding counter-diabatic terms that suppress the excitations \cite{Muga10,Junhua14}. 

In our scheme, similar to the scheme of adiabatic passage, we sweep the amplitude $\Omega (t) = \Omega_{0} \left[\sin(\pi t/T) + i\beta \right]$ and the detuning $\Delta(t)=\Delta_{0} \cos(\pi t/T)$ of the experimental control parameters in $H_{\rm aJC}$, where $\Omega_{0} = (2 \pi) 38.5$ kHz, $\beta = 0.075$, and $\Delta_{0} = 1.6 \Omega_{0}$ as shown in Fig. \ref{fig1:concept}(d). Note that we add the counter-diabatic phase control $i\beta$, which makes the state of the system remain in the instantaneous ground state during the fast driving \cite{Muga10,Morsch12,Junhua14}. On top of these controls, we add the compensation term for the AC-stark shift due to the coupling of the blue-sideband transition to the qubit level (see Appendix A). For the cancellation of the dynamic phases differently acquired during the population transfer, we invert the sign of $\Omega$ and reverse the control of $\Delta$ in the middle of the sequence. The total duration of $T$ for the transfer is 7 times longer than the $\pi$-pulse time of dynamic blue-sideband transition on $\ket{n=0}$. Note that the standard adiabatic scheme is about 20 times longer than the $\pi$-pulse duration for the same fidelity \cite{Junhua14} and a rapid adiabatic passage needs a coupling strength four times stronger than $\Omega_{0}$ for the same transfer efficiency \cite{Wunder07}. Fig. \ref{fig1:concept}(f) clearly shows the uniform transfer of population from $\ket{\downarrow, n}$ to $\ket{\uparrow, n+1}$ in the range from $n=0$ to 5.  

\begin{figure}[ht]
 \includegraphics[width=1.0\columnwidth]{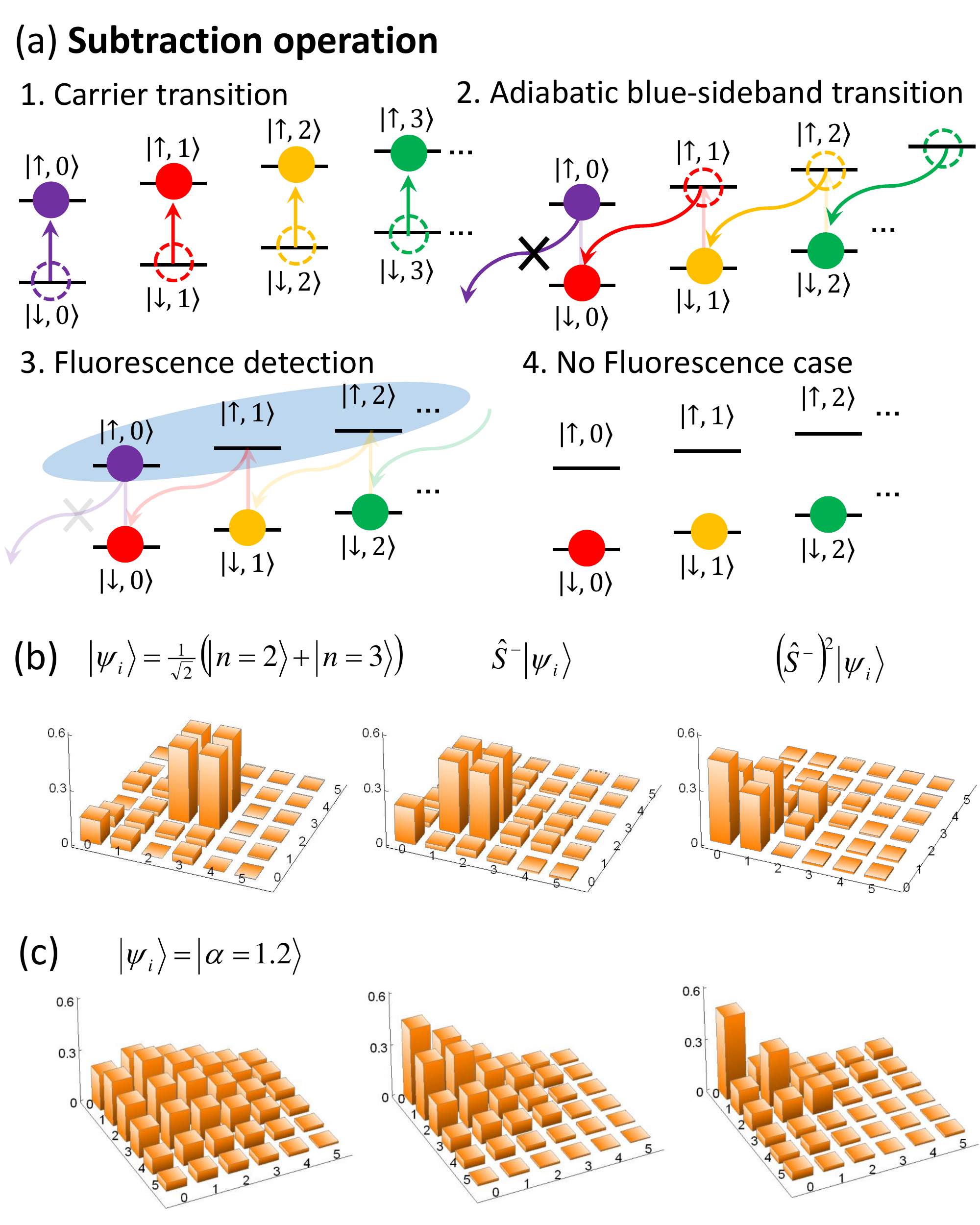}\\
\caption{Schematic diagram and experimental results of phonon subtraction. (a) Sequence of subtraction operations: the sequence of the operations for addition is reversed, $i. e.$,  a $\pi$-pulse of carrier transition followed by a $\pi$-pulse of adiabatic blue-sideband transition, which implements $\sum_{n} \ket{\downarrow, n} \bra{\uparrow, n+1}$. This takes the phonon state from $\ket{n+1}$ to $\ket{n}$ except $\ket{n=0} $ where $\ket{\downarrow, 0}$ transfers to $\ket{\uparrow, 0}$. The $\ket{\uparrow, 0}$ state is abandoned by the conditional measurement after the fluorescence detection which only collects non-zero phonon state data without fluorescence. (b) Subtraction on a superposition state $\ket{\psi_{i}}=\frac{1}{\sqrt{2}}\left( \ket{n=2}+\ket{n=3}\right)$. The population is reduced and the coherence is conserved. The initial fidelity and purity of the state are 0.83(0.02) and 0.73(0.03). The fidelities are changed to 0.77(0.02) and 0.83(0.01) after one and two times subtraction, respectively. The purities become 0.65(0.02) and 0.75(0.02). In the preparation and displacement operations for the superposition states with fluorescent detection, the zero components are increased due to unexpected experimental imperfections, which accidentally increase the fidelity and purity for the state $\frac{1}{\sqrt{2}}(\ket{0}+\ket{1})$. (c) Subtraction from an initial coherent state $\ket{\alpha=1.2}$. The initial fidelity of the state is 0.96(0.01) and the fidelities are reduced to 0.92(0.01) and 0.66(0.01) after one and two times subtraction, respectively.}
\label{fig3:subtraction}
\end{figure}

\begin{figure*}[ht]
\includegraphics[width=1.4\columnwidth]{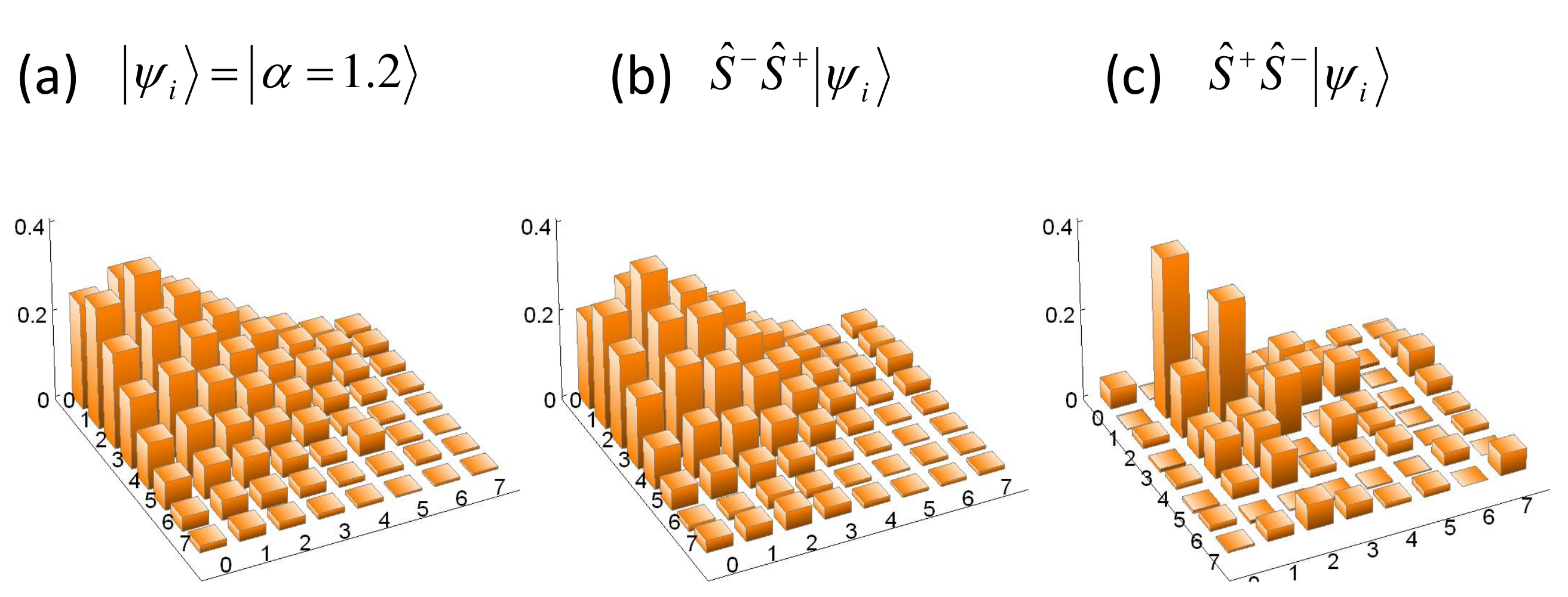}\\
 \caption{Experimental results after addition-then-subtraction and subtraction-then-addition, respectively. Only the real part of experimentally measured density matrices is shown $\bf{a.}$ for an initial coherent state $\ket{\psi_{i}}=\ket{\alpha=1.2}$, $\bf{b.}$ single-phonon added-then-subtracted $\hat{S}^{-}\hat{S}^{+}\ket{\psi_{i}}$ state and $\bf{c.}$ single-phonon subtracted-then-added $\hat{S}^{+}\hat{S}^{-}\ket{\psi_{i}}$ state. $\bf{b.}$ The state after addition-then-subtraction is basically identical to the original state. The fidelity of the $\hat{S}^{-}\hat{S}^{+}\ket{\psi_{i}}$ state to the original state $\ket{\psi_{i}}$ is 0.97(0.01) and the purity is 0.96(0.01). $\bf{c.}$ The state after subtraction-then-addition is not same as the original state, because the vacuum component is thrown away during the projective measurement. The small population in zero component mainly comes from the imperfection of the fluorescence detection and heating of the system, which is in good agreement with numerical simulation.}
\label{fig4:addsub}
\end{figure*}

\section{III. Conventional Addition Operation}
We implement the addition operation $\hat{S}^{+}$ in Eq. (\ref{eq:addsubtract}) by first applying the uniform blue-sideband transfer $\sum_{n=0} \ket{\uparrow, n+1} \bra{\downarrow, n} + \rm{h.c.}$ and then $\pi$-pulse of carrier transition 
$\sum_{n=0} \ket{\downarrow, n} \bra{\uparrow, n}$ $+\rm{h.c.}$ as shown in Fig. \ref{fig2:addition}(a). Our addition scheme deterministically adds one phonon independent of the initial phonon number state. We observe that quantum coherence is preserved in the addition operations. We prepare an initial state $\frac{1}{\sqrt{2}}\left( \ket{n=0}+\ket{n=1}\right)$, apply the additions up to three times and measure the density matrix of the resulting phonon states. As shown in Fig. \ref{fig2:addition}(b), the coherences represented by the off-diagonal terms of the density matrix clearly remain after addition processes. We construct the density matrix by using the iterative maximum-likelihood algorithm \cite{Jezek01} after displacing the state by 8 different angles with the amplitude of $\alpha \sim 0.8$ and then by finding the phonon number distributions (see Appendix B). The phonon number distributions are obtained by observing the time evolutions of the standard blue-sideband transitions, similar to the direct reconstruction scheme of the phonon density matrix \cite{Leibfried96}.

As a second example, we prepare an initial coherent state $\ket{\alpha=0.81}$ and apply the addition operations as shown in Figs. \ref{fig2:addition}(c) and (d). One immediate consequence of the addition operations on the coherent state is the production of sub-Poissonian phonon statistics because the addition increases the average phonon number but not the shape of the distribution and variance. We observe that the ratios between the variance and the average phonon number reduce from 1 to 0.43, 0.39, 0.2, after one to three single-phonon additions, respectively. Applying the first addition operation, we detect negativity in the Wigner function as shown in the second column of Fig. \ref{fig2:addition}(d). It is important to note that the addition operation, which converts a coherent state to a highly non-Gaussian state, is nearly deterministic unlike the case of $\hat{a}^{\dagger}$ operation \cite{Bellini07,Bellini08,Bellini09}. There is a limit in the number of additions we can apply, due to the validity of the adiabatic approximation and the heating process of phonons \cite{Myatt00}. Under this limitation, we could perform the operations three times without the significant loss of fidelity.  As shown in Fig. \ref{fig2:addition}(d), the experimental results and the theoretical predictions for the Wigner functions are in excellent agreement. This is significant in comparison to the photonic realization of bosonic operations of single photon creation and annihilation \cite{Bellini09}. Here we obtain the Wigner function of the state from the reconstructed density matrix.

\section{IV. Conventional Subtraction Operation}
The subtraction operation $\hat{S}^{-}$ in Eq. (\ref{eq:addsubtract}) is realized by reversing the sequence of the addition operation, that is, the application of the $\pi$-pulse of carrier transition and the uniform blue-sideband transfer $\sum_{n=1} \ket{\downarrow, n-1} \bra{\uparrow, n}+{\rm h.c.}$, followed by the fluorescent detection as shown in Fig. \ref{fig3:subtraction}(a). The zero phonon state $\ket{n=0}$ is eliminated after the subtraction, which is implemented by the conditional measurement in our experimental scheme. After the detection sequence, we only collect the data with no fluorescence, which has the success rate given by the probability of the non-zero phonon states. We examine the performance of the subtraction operation with an initial phonon superposition state $\frac{1}{\sqrt{2}}\left( \ket{n=2} + \ket{n=3} \right)$. As shown in Fig. \ref{fig3:subtraction}(b), the subtraction operation reduces the phonon excitation by one quanta. After the second application of the subtraction, the off-diagonal terms of the density matrix are significantly reduced, which shows the current limit in experiments due to the heating of the system. We also prepare a coherence state $\ket{\alpha=1.2}$ and apply the subtraction twice as shown in Fig \ref{fig3:subtraction}(c), which shows that qualitatively the subtraction works for any initial quantum state. The subtraction operation can squeeze a coherent state which is different from annihilation that has the coherence state as its eigenstate. However, our experimental precision is not high enough to observe the squeezing effect.

\section{V. Commutation Relation of the Addition and the Subtraction}
We study experimentally how the quantum states are changed depending on the order of the addition and subtraction for an initial coherent state $\ket{\psi_{i}}=\ket{\alpha=1.2}$. If we add then subtract $\hat{S}^{-} \hat{S}^{+} \ket{\psi_{i}}$, the state after the sequence is the same as the original one, since there is no amplitude modification. For the case of subtraction-then-addition $\hat{S}^{+} \hat{S}^{-} \ket{\psi_{i}}$, the final state does not have vacuum component because the vacuum state is removed at the first subtraction. Fig. \ref{fig4:addsub}(b) shows the experimental result of $\hat{S}^{-} \hat{S}^{+} \ket{\psi_{i}}$, which is basically identical to the initial state of Fig. \ref{fig4:addsub}(a). Fig. \ref{fig4:addsub}(c) shows the result after the operation of $\hat{S}^{+} \hat{S}^{-} \ket{\psi_{i}}$, where there is no significant vacuum component in the density matrix. The vacuum component is not perfectly removed because of the detection error during the projective measurement based on the atomic fluorescence and heating of the system. The fluorescent detection duration is comparable to the motional coherence time of our system, which makes the off-diagonal part of the final state suppressed significantly (see Appendix C). Our experimental result is well in line with the non commuting relation of the Susskind-Glogower's phase operators, i.e., $[\hat{S}^{-},\hat{S}^{+}]=\ket{0}\bra{0}$ \cite{Glogower1964}.

\section{VI. Remarks}
Due to the capability of near-deterministically generating a non-Gaussian state, the conventional addition and subtraction operations provide a new efficient scheme for quantum state engineering \cite{Braunstein99,Braunstein05}. This scheme may be further applied to various quantum optics setups such as cavity quantum electrodynamics and optomechanics. We note a theoretical scheme that the non-Gaussian information of a phonon state would be transferred to that of photon through phonon-photon coupling \cite{Orszag02,Lamata11}. It has also been discussed that the conventional arithmetic addition and subtraction can be used to measure the vacuum state without disturbing the state \cite{Jeffers13} and therefore it can be directly applied to construct the Q-function of a quantum state of motion.

\section*{Acknowledgement}
This work was supported by the National Basic Research Program of China under Grants No. 2011CBA00300 (No. 2011CBA00301), the National Natural Science Foundation of China 11374178. K.Kim acknowledges the first recruitment program of global youth experts of China. M.S. Kim was supported by the UK EPSRC and Royal Society Wolfson Merit Award. H. Nha was supported by an NPRP grant 4-520-1-083 from Qatar National Research Fund.

\hspace{-0.55cm}$*$ m.kim@imperial.ac.uk\\
$^\dag$ kimkihwan@gmail.com

\section*{Appendix A: AC Stark shift compensation}
The AC Stark shift in the adiabatic operations mainly come from the off-resonant coupling to the carrier transition, the transition between $S_{1/2}\leftrightarrow P_{1/2}$ states of \Yb ion, and the other radial motional mode ($\omega_{\rm Y}\approx \omega_{\rm X}$ + $(2\pi)0.4$ MHz). The dominant AC stark shift comes from the carrier transition, which is $\frac{\Omega_{\rm 0}^2}{2 \omega_{\rm X}\eta^2} \sim (2 \pi) 33$ kHz, where $\Omega_{0} = (2 \pi) 38.5$ kHz and the Lamb-Dicke parameter $\eta=0.89$. The amount of the shift brought by Y mode is given by $\frac{\Omega_{\rm 0\rm Y}^2}{2 (\omega_{\rm Y}-\omega_{\rm X})}$ that is about 20 times smaller than that from the carrier coupling. The AC stark shift between qubit states from the Raman laser beams due to $S_{1/2}\leftrightarrow P_{1/2}$ transition is $\frac{g_{R1}^2+g_{R2}^2} {2\Delta_{R}} \frac{\omega_{\rm HF}}{\Delta_{R}} \sim (2 \pi) 1 $ kHz, where $g_{R1}$ and $g_{R2}$ are the coupling strengths of Raman 1 and Raman 2 beam \cite{Hayes10}, respectively and the detuning from $^2{P_{1/2}}$, $\Delta_{R}=(2\pi)$ 18 THz. 

We consider total AC stark shift as the form of $\frac{\left|\Omega(t)\right|^2}{2\Delta_{total}}$, and calibrate the blue sideband frequency to be $\omega_{bsb}^{real}=\omega_{bsb}^{meas}-\frac{\left|\Omega(t)\right|^2} {2\Delta_{total}}$, here $\omega_{bsb}^{meas}$ is the measured blue sideband frequency. We measure $\Delta_{total}$ by varying the intensity of the laser to get several sets of $\{\omega_{bsb}^{meas}$, $\Omega_{bsb}\}$ and by fitting the results with $\omega_{bsb}^{meas}=\omega_{bsb}^{real} +\frac{\Omega_{bsb}^2}{2\Delta_{total}}$. Then we apply the time dependent phase based on the amplitude $\Omega (t) = \Omega_{0} \left[\sin(\pi t/T) + i\beta \right]$ and the detuning $\Delta(t)=\Delta_{0} \cos(\pi t/T)$ as follows, 
\begin{widetext}
\begin{eqnarray}
\Omega_{0} \left[\sin(\pi t/T)\cos\left[\int_{0}^{t} (\omega_{bsb}^{real}+\Delta(t'))dt'\right] - \beta \sin\left[\int_{0}^{t} (\omega_{bsb}^{real}+\Delta(t'))dt'\right] \right].
\label{eq:RAPformula}
\end{eqnarray}
\end{widetext}
Note that the imaginary part here implies $\frac{\pi}{2}$ phase difference. Here, $\int_{0}^{t} (\omega_{bsb}^{real}+\Delta(t'))dt'$ is calculated as follows based on the measurements of $\omega_{bsb}^{meas}$ and $\Delta_{total}$,

\begin{widetext}
\begin{eqnarray}
\int_{0}^{t} \left[\omega_{bsb}^{real} + \Delta(t') \right]dt'&=&
\int_{0}^{t} \left[\omega_{bsb}^{meas}-\frac{\left|\Omega(t)\right|^2} {2\Delta_{total}}+\Delta_{0} \cos(\pi t/T)\right]dt' \\
&=& \omega_{bsb}^{meas} t - \frac{\Omega_{\rm 0}^2} {2\Delta_{total}} \int_{0}^{t} \left[ \sin^2(\pi t/T) + \beta^2 \right] dt' + \Delta_{0} \frac{T}{\pi} \sin(\pi t/T)\\
&=& \omega_{bsb}^{meas} t - \frac{\Omega_{\rm 0}^2} {4\Delta_{total}} \left[ (1+2\beta^2)t + \frac{T}{2\pi} \sin(2\pi t/T) \right] + \Delta_{0} \frac{T}{\pi} \sin(\pi t/T).
\label{eq:PAPphase}
\end{eqnarray}
\end{widetext}

\section*{Appendix B: Reconstruction of phonon density matrix}
We use an iterative algorithm proposed in Ref. \cite{Jezek01} for the reconstruction of an unknown state. It consists of a maximum-likelihood estimation solved by expectation-maximization algorithm followed by a unitary transformation of the eigenbasis of the density matrix $\rho$.

Based on the measured phonon distribution $f_{n}$ of $N$ measurements by different displacement, we aim to get real probabilities $p_{n}=\bra{n}\rho\ket{n}$ that are as close to the observed frequencies $f_{n}$ as possible, which can be subject to the maximum-likelihood functional
\begin{eqnarray}
\ln L\left(\rho\right)=\ln \prod\limits_n\bra{n}\rho\ket{n}^{f_{n}} =-\sum_{n}f_{n}\ln p_{n},
\label{eq:ml}
\end{eqnarray}
from which we reconstruct $\rho$. This likelihood functional can be interpreted as a linear and positive (LP) problem in the classical signal processing:
\begin{eqnarray}
p_{n}=\sum_{i}r_{i}h_{in},
\label{eq:lprelation}
\end{eqnarray}
where $r_{i}$ are eigenvalues of $\rho$\ and $h_{in}$ is a positive kernel. We can solve this LP problem with the expectation-maximization algorithm \cite{dempster1977maximum,vardi1993image}:
\begin{eqnarray}
r_{i}^{\left(k\right)}=r_{i}^{\left(k-1\right)}
\sum_{n}\frac{h_{in}f_{n}}{p_{n}\left(\bf{r}^{\left(k-1\right)}\right)},
\label{eq:iteration}
\end{eqnarray}
which is initially set to a positive vector $\bf{r}$ $\left(r_{i}>0 \; \forall i\right)$.

The second part aims at getting the eigenbasis diagonalizing the density matrix. This part consists of two steps: reconstruction of the eigenvectors of $\rho$ in a fixed basis, and rotation of the basis using a unitary transformation
\begin{eqnarray}
\ket{\phi_{n}^{'}}\bra{\phi_{n}^{'}}=U\ket{\phi_{n}}\bra{\phi_{n}}U^{\dagger}
\label{eq:unitrans}
\end{eqnarray}
with the infinitesimal form $U\equiv e^{i\epsilon G} \approx 1+i\epsilon G$ and $\epsilon$ is a small positive real number. $G=i\left[\rho,R\right]$ is chosen as a Hermitian generator of the unitary transformation, where $R$ is a semipositive definite Hermitian operator $R=\sum_{n}\frac{f_{n}}{p_{n}}\ket{n}\bra{n}$.

Starting from some positive initial density matrix $\rho$, we continue repetition of first finding new eigenvalues $r_{i}$ using the expectation-maximization iterative algorithm (\ref{eq:iteration}) and then finding new eigenvectors ${\phi_{i}}$ by unitarily transforming the old ones. The likelihood of the estimate $p_{n}$ is increased and we finally reach to determine the density matrix
\begin{eqnarray}
\rho=\sum_{n}r_{n}\ket{\phi_{n}}\bra{\phi_{n}}.
\label{eq:parameterize}
\end{eqnarray}

\section*{Appendix C: Error Analysis}
Dominant error comes from phonon heating process caused by the electric-field noise from the trap electrodes \cite{Turchette00}. Heating decreases the Fock state preparation fidelity and affects the adiabatic blue sideband process. Its time evolution is known to be described by \cite{Turchette00,Wineland00}:
\begin{widetext}
\begin{eqnarray}
\dot{\rho}(t)=\frac{\gamma}{2}\bar{n}\left(2\hat{a}^{\dagger}\rho\left(t\right)\hat{a}
-\rho\left(t\right)\hat{a}\hat{a}^{\dagger}-\hat{a}\hat{a}^{\dagger}\rho\left(t\right)\right)
+\frac{\gamma}{2}\left(\bar{n}+1\right)\left(2\hat{a}\rho\left(t\right)\hat{a}^{\dagger}
-\rho\left(t\right)\hat{a}\hat{a}^{\dagger}-\hat{a}^{\dagger}\hat{a}\rho\left(t\right)\right).
\label{eq:heating}
\end{eqnarray}
\end{widetext}
Here $\gamma$ is the coupling strength between the ion motion and the thermal reservoir, $\bar{n}$ is the average phonon of the thermal reservoir. In our experimental setup, the heating rate $\gamma\bar{n}\approx\gamma\bar{n+1}$ is 150 Hz. It can be reduced by using a large trap, cleaning the electrodes \cite{hite2012100} (equivalent to reducing $\gamma$) or cooling the trap (equivalent to reducing $\bar{n}$).

More error in experiments may be caused by fitting since some noise in the blue-sideband curves may be recognized as a high phonon population. Although the Fock state preparation error will be involved in the transition probability, fortunately the error for a small $n$ is relatively insignificant and the population mainly resides in small $n$ values for the initial states we prepared in the experiments.

There are some other small errors that we did not calculate or simulate, including the fluctuation of the trap frequency $\sim$0.1 kHz coming from $\sim$2\% intensity fluctuation of Raman laser, $\sim$1 kHz trap frequency fluctuation coming from sudden changes of the ion position. We can see how much our experimental results are modified when comparing to the simulations \cite{Shuoming14}.







\end{document}